\pdfoutput=1
\documentclass[referee]{raa}            % referee version: for submission

\usepackage{graphicx,times}             %for PS/EPS graphics inclusion, new
\usepackage{natbib}
\usepackage{amssymb,amsmath}
\usepackage{color}
\bibpunct{(}{)}{;}{a}{}{,}
\usepackage[a4paper=true,pagebackref=true]{hyperref}
\hypersetup{colorlinks = true, linkcolor = green, anchorcolor = red, citecolor = blue, filecolor = red, pagecolor = red, urlcolor = red}

\begin{document}

   \title{A Method for Pulsar Searching: Combining a Two-dimensional Autocorrelation Profile Map and a Deep Convolutional Neural Network
%\footnotetext{$*$ Supported by the National Natural Science Foundation of China.}
}
%   \subtitle{I. Place Your Subtitle Here}

   \volnopage{Vol.0 (20xx) No.0, 000--000}      %%preserved for Editor. DOn't remove!
   \setcounter{page}{1}          %%starting page, preserved for Editor. DOn't remove!

   \author{Long-Qi Wang
      \inst{1}
   \and Jing Jin
      \inst{1}
   \and Lu Liu
      \inst{1}
   \and Yi Shen
   \inst{1}
   }

   \institute{Harbin Institute of Technology,
   	Department of Control Science and Engineering,
   	Room 502, Main Building, 92 West Dazhi Street,
   	Nangang District, Harbin 150001, China; {\it jinjinghit@hit.edu.cn}\\
\vs\no
   {\small Received~~20xx month day; accepted~~20xx~~month day}}

\abstract{ In pulsar astronomy, detecting effective pulsar signals among numerous pulsar candidates is an important research topic. Starting from space X-ray pulsar signals, the two-dimensional autocorrelation profile map (2D-APM) feature modelling method, which utilizes epoch folding of the autocorrelation function of X-ray signals and expands the time-domain information of the periodic axis, is proposed. A uniform setting criterion regarding the time resolution of the periodic axis addresses pulsar signals without any prior information. Compared with the traditional profile, the model has a strong anti-noise ability, a greater abundance of information and consistent characteristics. The new feature is simulated with double Gaussian components, and the characteristic distribution of the model is revealed to be closely related to the distance between the double peaks of the profile. Next, a deep convolutional neural network (DCNN) is built, named Inception-ResNet. According to the order of the peak separation and number of arriving photons, 30 data sets based on the Poisson process are simulated to construct the training set, and the observation data of PSRs B0531+21, B0540-69 and B1509-58 from the Rossi X-ray Timing Explorer (RXTE) are selected to generate the test set. The number of training sets and the number of test sets are 30,000 and 5,400, respectively. After achieving convergence stability, more than 99$\%$ of the pulsar signals are recognized, and more than 99$\%$ of the interference is successfully rejected, which verifies the high degree of agreement between the network and the feature model and the high potential of the proposed method in searching for pulsars.
\keywords{methods: data analysis; methods: statistical; X-rays: stars}
}

   \authorrunning{L. Wang, J. Jin, L. Liu \& Y. Shen }            %author_head in even pages
   \titlerunning{Pulsar searching based on a new proposed method}  % title_head in odd pages

   \maketitle
%% The author head (on even pages) and the title head (on odd pages) will be
%% automatically extracted from \author{} and \title{}. Whenever the title is too long,
%% you will be asked to supply a shorter one by inserting either \authorrunning{} or
%% \titlerunning{} before \maketitle. Anyway, you can specify your own heads.
%%
%%
%% Note: In the following text body of your manuscript, please note several differences from
%%       other major journals:
%% (1) \subsection{Please Capitalize the First Letter of Each Notional Word in Subsection Title}
%% (2) Please Capitalize the First Letter of Each Notional Word in all tables' captions

%
%________________________________________________ sections below

% \textcolor{red}{\sout{caoninainai}}
\section{INTRODUCTION}           %% first-level sections will be auto-capitalized
\label{sect:intro}

A pulsar is a type of high-density neutron star with a strong magnetic field and high rotation speed. Discoveries related to pulsars have led to two successive Nobel prizes \citep{1968Nature...217...709,1975APJ...195...L51}. Pulsars can be divided into magnetars, accretion-powered pulsars and rotation-powered pulsars according to the energy sources of their electromagnetic radiation. The rotation-powered pulsars, which have a stable rotation period and fixed magnetic inclination, radiate cyclical signals and can be classified into normal pulsars and millisecond pulsars depending on the period range. Normal pulsars, the rotation periods of which vary from a few tens of milliseconds to several seconds, account for 90$\%$ of the total pulsar population and provide a large number of statistically significant samples for tasks such as the validation of the pulsar evolution model \citep{2006APJ...643...332} and exploration of the interstellar medium \citep{1993APJ...411...674}, statistics related to the birth rates of neutron stars in the galaxy \citep{2008MNRAS...391(4)...2009}, guidance for deep space exploration \citep{2006J...GUID...CONTROL...DYNAM...29...49, 2016EXP...ASTRON...42(2)..101}, etc.

More than 2,500 pulsars have been included in the Australia Telescope National Facility (ATNF); their parameters can be queried and visualized from the official ATNF website. Most of these pulsars were discovered by ground-based sky survey instruments, including the Parkes Multi-beam Pulsar Survey (PMPS) \citep{2001MNRAS...328(1)...17}, High Time Resolution Universe Survey (HTRU) \citep{2010MNRAS...409(2)...619}, Pulsar Arecibo L-band Feed Array Survey (PALFA) \citep{2009APJ...703(2)...2259}, LOFAR Tied-Array All-sky Survey (LOTAAS) \citep{2014AA...570...A60}, Greenbank Northern Celestial Cap Survey (GBNCC) \citep{2014APJ...791(1)...67}, Five Hundred Metre Aperture Spherical Telescope (FAST) \citep{2011IJMPD...20(6)...989}, and Square Kilometre Array (SKA) \citep{2009AA...493(3)...1161}.

However, the absorption of high-energy photons by the Earth's atmosphere limits the effectiveness of ground observations. The development of space detection technology has shown that the radiation energy of normal pulsars is concentrated in the X-ray region. Some pulsars show all the properties of radio pulsars observed at high energies but emit almost no radio pulses, such as the famous Geminga (named PSR J0633+1746 in the pulsar catalogues) \citep{2014APJ...793(2)...905}. \citet{2012CUP...250} mentioned that observation from space at X-ray energies promises to intuitively show the electromagnetic radiation processes rather than the view provided by radio observations, as at radio energies, observations suffer from interference by uncertain physical factors, including the emission coherence and propagation effects in the interstellar medium and pulsar magnetosphere. Using optical observations, thermal and non-thermal processes are difficult to distinguish from the available data. Compared with the vast areas covered by ground observations, the receiving area of a space detector is limited by the load and volume of the spacecraft, which often submerges the photon signals in an uncertain Poisson distribution. However, in the long run, space observations still have great potential and value. Numerous space exploration satellites, including Ariel5 \citep{1981APJ...243...155}, HEAO-1 \citep{1980APJ...235...377}, HEAO-2 \citep{1981IEEET...28(1)...869}, ROAST \citep{1999AA...349...389}, RXTE \citep{1998MNRAS...300(2)...583}, Chandra \citep{2017APJ...840(2)...94}, NuSTAR \citep{2016APJ...817(2)...93}, and HXMT \citep{2004ASR...34(12)...2667}, have been launched. Most of the received on-orbit data can be accessed from the National Aeronautics and Space Administration (NASA)'s High Energy Astrophysics Science Archive Research Center (HEASARC). However, screening valuable pulsar signals from this massive amount of data to enable further observations has become a critical problem.

Scholars have conducted considerable work on classifying pulsar candidates, but most of these studies are focused on radio astronomy. Automation, high accuracy and fast processing speed are common goals, and the screening methods have generally been developed with techniques that range from artificial recognition to machine learning algorithms. It is difficult for artificial recognition techniques to address the increasing amount of sample data, but the developed techniques include much useful wisdom, such as the information classification method based on the signal-to-noise ratio (S/N) \citep{1992MNRAS...255(3)...401, 1986APJ...311...694}, graphic software-aided classification \citep{2004MNRAS...355(1)...147,2009MNRAS...395(2)...837} and a software-based scoring system \citep{2013MNRAS...433(1)...688} based on more empirical features, including the pulse width, S/N of the pulse, dispersion measure value, period information, signal duration in the time and frequency domains, etc.

With the development of artificial intelligence techniques, \citet{2010MNRAS...407(4)...2443}, \citet{2012MNRAS...427(2)...1052} and \citet{2014MNRAS...443(2)...1651} introduced more empirical features as the input to an artificial neural network (ANN), which greatly improved the accuracy and processing speed of pulsar searches. After undergoing supervised training, the network acquired automatic classification skills and achieved satisfactory results for the PMPS and HTRU data sets. However, these empirical features may make the network over-reliant on experience and hypotheses while failing to consider other useful information. Moreover, these features manifest differently in different pulsars and different data sets, which increases the network training difficulty. In addition, expanding the number of sample sets to cover all types of situations as much as possible influences the final prediction accuracy. To overcome the tendency of feature dependence, \citet{2016MNRAS...459(1)...1104} extracted unbiased statistical features from the pulse profile and dispersion measure (DM) curves to form the input of a fast decision tree (FDT) and achieved notable results on the HTRU and LOTAAS data sets. \citet{2014APJ...781(2)...117} employed a convolutional neural network (CNN) to directly learn the characteristics of the profile and DM curves. This approach eliminated the need for artificial feature design and achieved high accuracy on the PALFA data sets.

Currently, algorithms are designed for data sets from different ground observation instruments. The number, signal quality, and positive and negative distributions of the sample data sets differ, but these algorithms provide reasonable guidance for our current research. We noticed that the pulsar candidate classification at space X-ray energies requires further study. Compared with ground radio observations, this type of classification does not need to consider the influence of the dispersion effect; thus, it naturally avoids the complex de-dispersion process. However, the statistical information in these data is sparse, and the DM curve is not meaningful. Moreover, the S/N in the observation data is lower. Therefore, a complete and reasonable pulsar searching algorithm at X-ray energies is urgently needed.

To quickly and accurately obtain an effective low S/N signal within a finite observation time from among massive number of candidates, we started from two aspects: feature representation and feature perception. Feature representation can be obtained by introducing autocorrelation and stacking the folding profiles across the trial periods. To address the pulsar signal with an unknown period and quality in practical applications, a time resolution setting criterion of the period axis is proposed to ensure the consistency of the extracted features. Although the feature modelling method is not complicated, the proposed feature improved the traditional time-domain profile in several parts. The method simultaneously enhances the feature consistency and preserves more details. On the other hand, the method suppresses the loss of useful information and resists shape distortions caused by the randomness of noise and insufficient accuracy of the folding period \citep{2016IEEETAES...52(5)...2210}, ultimately guaranteeing a more comprehensive representation of the essential signal characteristics. Regarding feature perception, we employ the modules from the famous Inception and ResNet CNNs to build a deep convolution neural network \citep{2015CVPR...1, 2016CVPR...2818, 2016CVPR...770, 2017AAAI} to learn the spatial information of the two-dimensional autocorrelation profile map (2D-APM). To train the network in a flexible and simple manner, a simulation method based on the non-homogeneous Poisson process is utilized to construct the training set. In the simulation process, only the peak separation and number of arrival photons need to be considered.
The network can complete the task of pulsar candidate identification precisely by virtue of its high sensitivity to the proposed model. The performance of the algorithm proposed in the paper has been verified and discussed in detail using Rossi X-ray Timing Explorer (RXTE) data.

The remainder of this paper is organized as follows: The proposed method , including feature extraction and representation, network structure building and the specific training strategies, is described in Section \ref{sec:Method}. Section \ref{sec:Observation} introduces the data preprocessing operations, including the training set generated by the simulation data and the test set generated by the actual observation data. In Section \ref{sec:results}, the results of the training and network classification ability are discussed and analysed. The conclusions are given in Section \ref{sec:conclusions}.

\section{METHODS} \label{sec:Method}

\subsection{2D-APM\label{subsec:2dapm}}

X-ray detection information includes the time of arrival (TOA) of X-ray photons. \citet{2011IEEETAES...47(4)...2317} modelled X-ray pulsar signals based on the non-homogeneous Poisson process (NHPP) and provided a mathematical description of the epoch-folding algorithm. The concept of epoch folding is easy to understand; the data stream is folded into different phase bins in each estimation period to obtain the possible profile information. The reciprocal of the period is the frequency, which makes it easy to consider the frequency domain. Each frequency point in the spectrum corresponds to a pulsar candidate—potentially a real pulsar signal. According to the periodic characteristics of the pulsar signal, the amplitude at the rotation frequency is generally larger. In practice, the folding profile in the time domain and S/N in the frequency domain are traditional effective features for confirming an X-ray pulsar signal.

However, the traditional features have some shortcomings. First, the X-ray photons are often submerged in the uncertainty of their Poisson distribution; correspondingly, the effective frequency points are submerged in the observation noise when the observation time is insufficient. Second, a small error in the folding period will lead to phase deviations or distortions of the profile shape \citep{2016IEEETAES...52(5)...2210}. However, note that when the folding period error falls within a reasonable range, the obtained profile still possesses some useful information, such as a slower change in the steep degree of the profile curve and the fixed direction of the phase shift. The other frequency points do not possess this property. If we can integrate the variation information of the trial periods based on their ground-truth values, the obtained features will be more abundant and will be able to suppress the adverse effects caused by random noise and estimated period inaccuracies.

\begin{figure}
	\centering
	\includegraphics[width=8cm,height=6.5cm]{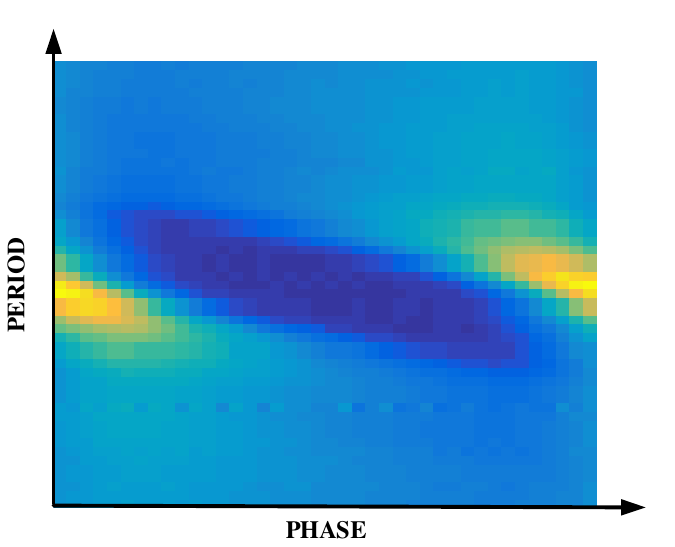}
	\caption{Schematic diagram of the 2D-APM, which can yield an important type of two-dimensional spatial distribution law after autocorrelation, epoch folding and time-domain feature expansion for X-ray light curves.}
	\label{fig:Schematic diagram of 2D-APM}
\end{figure}

Our goal is to enable an automatic and efficient search process. Currently, deep neural networks possess arbitrary non-linear expressiveness of complex models, which allow them to mine some information that the naked eye cannot perceive, especially from professional and non-daily aspects. 
The key step is to extract features that have strong regularity and more information for network learning. However, when the starting point of the folding signal is different, the final profile exhibits a phase shift. In pulsar candidate searching, the phase information is redundant, and different profile curves with different phases will undoubtedly increase the difficulty of network training and reduce the classification accuracy. Consequently, we apply a simple operation before folding. First, the time series of photon arrivals is transformed into discrete, equal-time-interval, real-valued light curves according to one certain sampling frequency, which represents the relationship between photon intensity and time. The precision of the light curve is equivalent to that of the time series when the sampling frequency is sufficiently large. Second, we perform light curve autocorrelation, which enhances the correlation of the effective signal and suppresses the uncorrelated noise, and on the other hand, ensures that the features of subsequent profiles are more consistent by virtue of its unbiased phase characteristics. The autocorrelation function reflects the correlation between the same light curves at different times, and
its mathematical form $R_{xx}(k)$ in the field of signal processing is expressed as follows:

\begin{equation}\label{equa2-1}
R_{xx}(k)=\lim\limits_{N\to\infty} \frac{1}{N}\sum_{n=0}^{N-1} x(n)x^{*}(n-k)
\end{equation}
This formula represents the discrete autocorrelation function of the discrete signal $x(n)$ at lag $k$. $N$ is the length of the observation signal, and $*$ represents the complex conjugate operation. This definition is often suitable for the ergodic stationary process or cyclostationary process and can give sensible well-defined single-parameter results for periodic functions. Here, the lag $k$ is essentially a concept of time, which is equivalent to the time delay information obtained by multiplying $k$ by the sampling time $T_s$. During a space mission, a detector may encounter abnormal situations, such as Earth occlusion, the South Atlantic Anomaly and electromagnetic contamination. In these cases, the received data are discontinuous; one advantage of autocorrelation is that it can directly combine these discontinuous data to enhance the overall S/N without requiring frequency correction or phase alignment.

As previously mentioned, the obtained features will be more abundant when the folding profile information can be integrated near the rotation period. The autocorrelation function can retain the periodic characteristics of the original signal; hence, the optimal folding period is still the rotation period of the pulsar. Therefore, we expand the time-domain features and propose the concept of a 2D-APM, as shown in Figure \ref{fig:Schematic diagram of 2D-APM}, where the horizontal axis represents phase bins with a normalized value of 0$\sim$1, the vertical axis represents the folding period, and the image brightness represents the number of photons obtained by epoch folding of the autocorrelation. After normalization of the maximum and minimum values, the brightness values can be adjusted from 0 to 255, which is consistent with the uint8 picture format. The traditional profiles represent the pulse information that is recorded when the emission beam sweeps the detector in each rotation. The number of pulses is usually one or two, and the pulses can be simulated with double Gaussian components. When the starting time for receiving photon signals varies, the initial phase of the traditional profile will change, and the overall waveform will shift. However, the autocorrelation profile and the 2D-APM will not change due to their phase-unbiased characteristics.

\begin{figure}
	\centering
	\includegraphics[width=16cm,height=10cm]{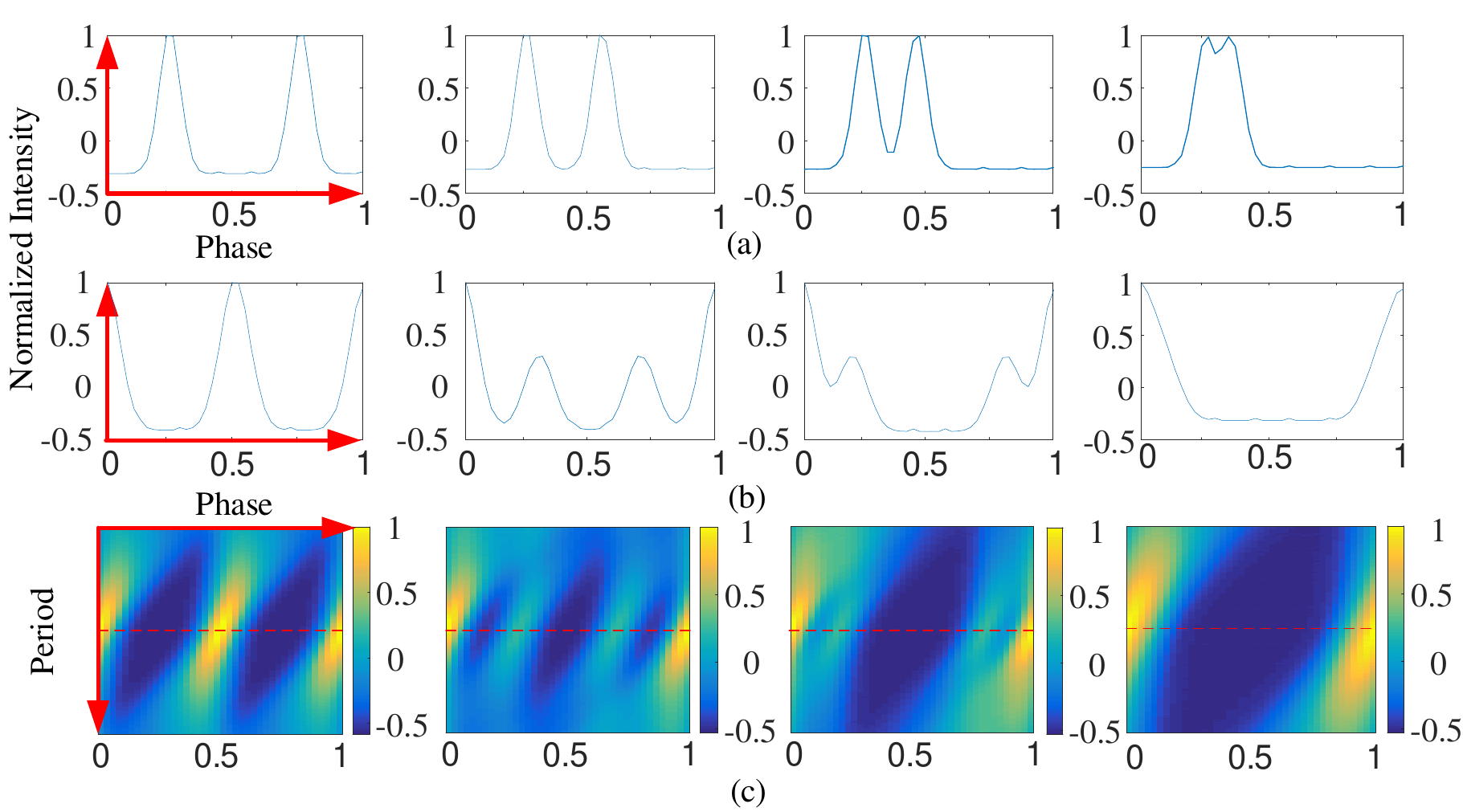}
	\caption{Pulse profiles (a), autocorrelation profiles (b) and 2D-APM (c) of different X-ray pulsar signals. Here, the sampling frequency is 1,000 Hz, the signal length is 40,000 (i.e., 10 s), and the rotation period is set to 33.65 ms. The maximum amplitude is uniformly 1 after normalization. The number of phase bins is 40 and normalized to 0$\sim$1. In (c), the entire two-dimensional matrix is normalized. The centre of the periodic axis is the accurate rotation period value, the time resolution is 0.1 us and the number of expansions is 299. Thus, the size of the 2D-APM is 299$\times$40. The red-dotted line at the centre of the periodic axis indicates the autocorrelation profile. }
	\label{fig:Simulation 1 of 2D-APM}
\end{figure}

Next, some simulation experiments are performed to illustrate the specific features of the 2D-APM. The number of arriving photons obeys a Poisson distribution; however, the simulation model has an additive mixed form that consists of the profile's periodic extension and Gaussian white noise \citep{2012CUP...126}. In applications where the count rates are high, the Gaussian approximation of the Poisson distribution is sufficiently accurate. For weak emission, the number of arriving photons per unit time is small, and the Poisson distribution is a discretely skewed distribution, which means that our model becomes less precise. However, the noise obtained by removing the profile signal is still consistent with the characteristics of a time-independent and constant power spectrum. We argue that using other (Gaussian) forms of white noise to replace it will not affect the final analysis results. Therefore, for convenience in simulation and analysis, we use a mixed form that consists of profile signals and Gaussian white noise to simulate the pulsar signals.

\begin{figure}[htbp!]
	\centering
	\includegraphics[width=11cm,height=8cm]{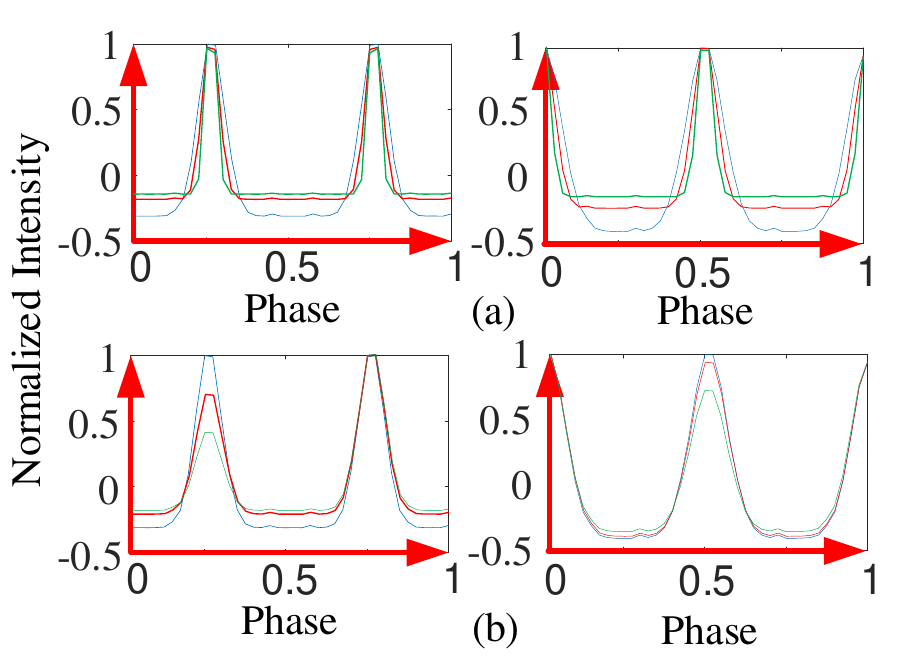}
	\caption{Result of the pulse width effect on the autocorrelation profile (a), and the result of the effect of the amplitude ratios of the double peaks on the autocorrelation profile (b). On the left is the pulsar profile curve with different shapes, and on the right is the corresponding autocorrelation profile curve plotted in the same line colour.}
	\label{fig:Simulation 2 of 2D-APM}
\end{figure}

The main parameters of the pulsar profile include the peak separation, pulse width and amplitude ratio of the double peaks. Figure \ref{fig:Simulation 1 of 2D-APM} shows the autocorrelation profile and the 2D-APM of X-ray pulsar signals with different peak separations. As Figure \ref{fig:Simulation 1 of 2D-APM} shows, there is only one peak at the centre of the autocorrelation profile when the two Gaussian components are far apart. As the two Gaussian components become closer, the peak spreads to both sides, and the profile eventually mutates into a U-shape. The autocorrelation profile, which is unlimited by the initial phase, has two peaks on both sides, at the points where the signal autocorrelation is strongest. 
As a result of stacking the autocorrelation profile near the rotation period, the 2D-APMs show consistent regularity with the autocorrelation profile and can provide more abundant information. Figure \ref{fig:Simulation 2 of 2D-APM} shows the effect extent of the pulse width and amplitude ratios on the autocorrelation profile. Clearly, the width of the autocorrelation profile is controlled by the pulse width, and the convex degree of the autocorrelation profile is controlled by the amplitude ratios. Compared with Figure \ref{fig:Simulation 1 of 2D-APM}(b), the factor that has the greatest effect on the autocorrelation profile or 2D-APM is the peak separation.

\begin{figure}
	\centering
	\includegraphics[width=14.5cm,height=6.5cm]{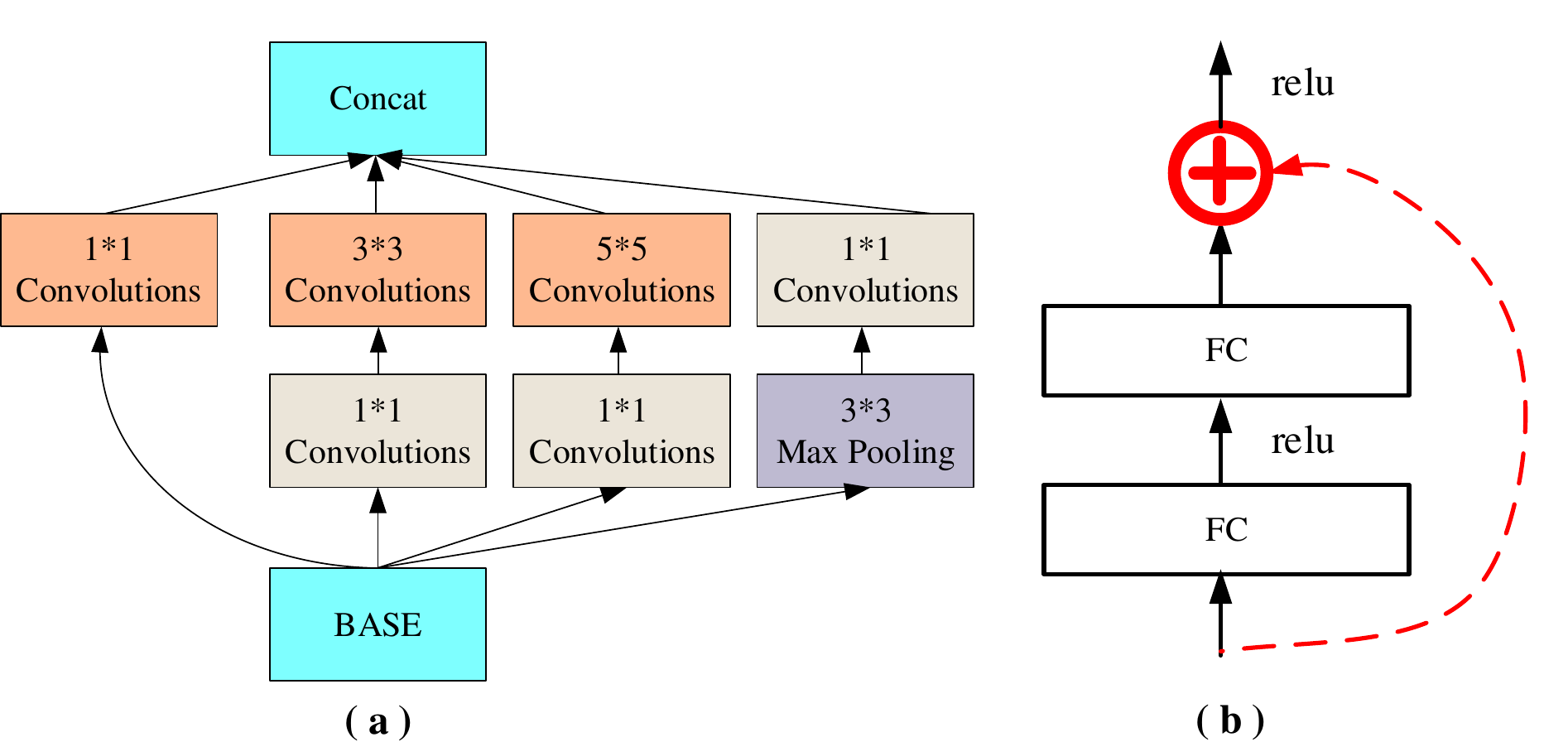}
	\caption{ Structure of the inception module in (a) and the basic residual learning network in (b). The inception is a core module of GoogleLeNet, and Base, as the basic input of the current inception module, can be either a Pooling layer or a Concat layer that connects the previous layer. The Concat layer merges the output matrices of the previous four layers into one matrix. FC represents the fully connected layer, and the rectified linear unit (ReLU) represents one type of activation function.}
	\label{fig:Resnet}
\end{figure}

\subsection{The Deep Convolutional Network \label{subsec:DCNN}}
%% Authors can give a citation as 'Michel et al. 1992'.
%% You may also use \cite, \citep and \citet for citation, and use Table~1 or Figure~1
%% and so forth. Using \ref and \label for cross-references of Tables/Figures
%% is a good way in adjusting/adding/removing text, tables or figures.

Deep learning, similar to the other machine learning algorithms, mainly acts as a feature learner to ensure that the training error converges without over-fitting. After the training error converges, the final effect depends on the quality of the feature modelling, such as the 2D-APM in this paper. Therefore, in practical application, a more targeted network can be built according to the actual requirements of the computing resources and computing speed. Here, the extracted feature is a two-dimensional matrix whose form is equivalent to that of the image; therefore, we selected the deep CNN (DCNN) named Inception-ResNet to model it. 
% \textcolor{red}{\sout{caoninainai}}  2016CVPR...2818

To use fewer computing resources to learn the feature information of the training sets in a shorter time, the network mainly uses the following skills. An inception module, shown in Figure \ref{fig:Resnet} (a), can achieve feature fusion at different scales by concurrently splicing the pooling layers and convolutional layers of different sizes \citep{2015CVPR...1}. The batch normalization (BN) structure, which ensures that the numerical distribution of each layer's input is as consistent as possible, largely solves the notorious exploding and vanishing gradient problems, accelerates convergence and improves the generalizability of the network model. \citet{2016CVPR...2818} further analysed and discussed the Inception module and customized a series of structure optimization rules to guarantee the most effective use of computing resources. These optimizations include dimensional reduction using a 1$\times$1 convolutional layer to realize lossless feature compression, the decomposition of large-scale convolutional layers into multiple small-scale convolutional layers in series, the insertion of 1$\times$N and N$\times$1 convolutional layers in series to replace the N$\times$N convolutional layers in the middle of the network, etc. ResNet, shown in Figure \ref{fig:Resnet} (b), can learn the mapping from the input to the residual. In terms of the final effect, ResNet offers a breakthrough by solving the convergence problem of deep networks \citep{2016CVPR...770}. A dropout layer is added before the final output layer. In the training process, this layer can deactivate some nodes in the current layer, that is, the output of these nodes is 0 to ensure the generalization ability of the network.

\begin{figure}[htbp!]
	\centering
	\includegraphics[width=8cm,height=14cm]{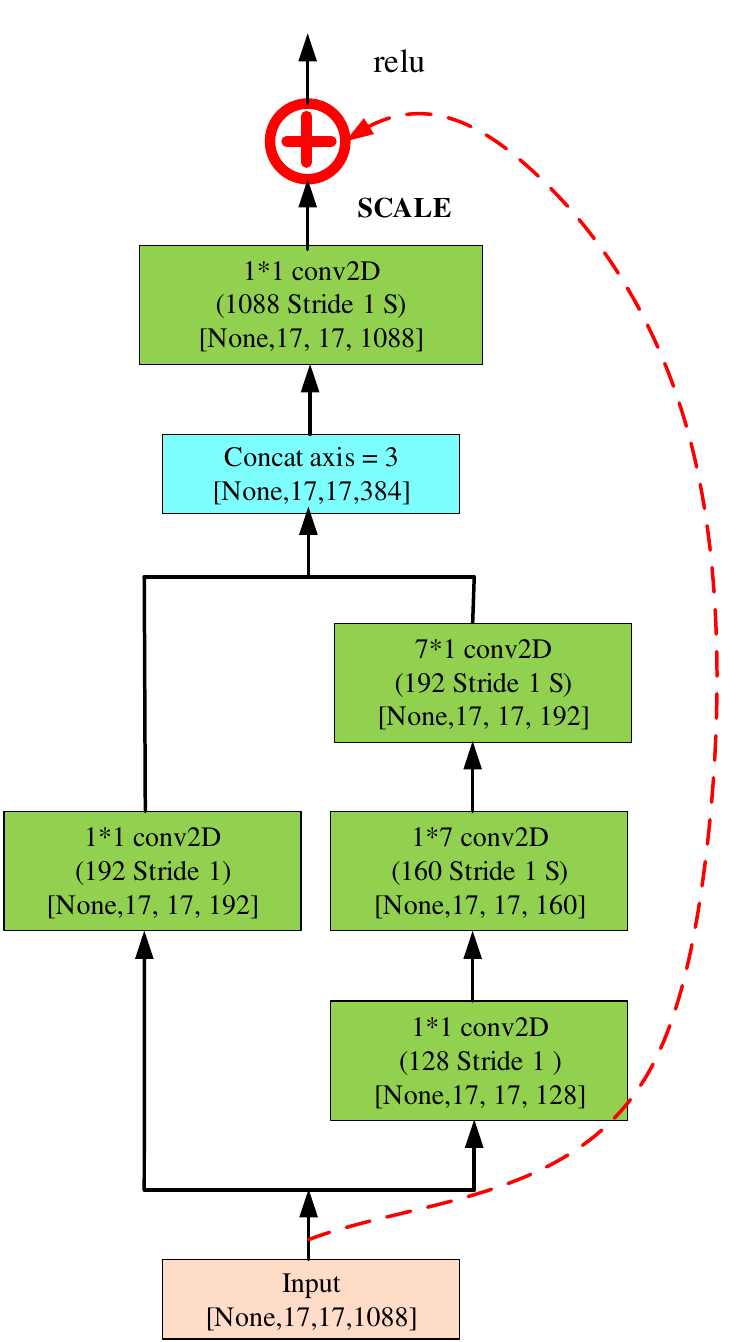}
	\caption{An important component of Inception-ResNet. The input and output size are both 17$\times$17, and the module is referred to as Block17. The green box represents a convolutional layer; the first line shows the size of the convolutional kernel, the second line shows the number of output channels and the step size of the traverse and filling modes, and the third line records the output size of this layer. The blue box represents the splicing layer, which splices the information on the fourth dimension of all the inputs. SCALE represents the shrinkage scale, which is usually between 0.1 and 0.3.}
	\label{fig:Inception-Resnet}
\end{figure}

As shown in Figure \ref{fig:Inception-Resnet}, Block17, which is a combination of one kind of inception basic configuration and the residual module (marked in red), is an important component of Inception-ResNet \citep{2017AAAI}. The number 17 means that the height and width of the data are always 17. Note that the input and output of the block module have the same size; thus, the depth of the network can be increased by connecting multiple identical block modules in series. In addition to the common FC layers and basic Inception module, the whole network is composed of multiple Block35s, Block17s and Block8s. Block35, Block17, and Block8 are repeated in the original Inception-ResNet network 10 times, 20 times, and 9 times, respectively. Their repeat times are adjusted to 3 in this paper to reduce the network size to match our research problem. The parameters can be adjustable, when facing different data sizes.

The input of the network consists of uint8 pictures with a size of 299$\times$299. In Section \ref{subsec:2dapm}, we mentioned that the size of the extracted 2D-APM is 299$\times$40, which is adjusted to 299$\times$299 by linear interpolation. One training iteration consists of one forward propagation and one backward propagation. The batch size, which is the number of samples utilized during a training process, is set to 20. The output of the network is simple two-dimensional information that signifies the probabilities of `yes' or `no'. Thus, the label can easily be changed from [0.0, 1.0] to [1.0, 0.0]. Both the cross-entropy between the output and the label and the regularization terms regarding the weighting parameters of the convolutional layers are simultaneously added to the loss function, which further reduces the possibility of overfitting and accelerates convergence. Regarding the training strategy of the network, we selected the momentum optimizer to update the training parameters. This optimizer reduces the oscillation trend of training and accelerates the convergence process while increasing the network's ability to escape local optima. The different available optimizers have advantages and disadvantages. The optimizer selected for this study performs well and can quickly find the optimal model solution.

The network is built and realized based on a TensorFlow framework. The software configuration includes CUDA 9.0.176-win10, cudnn-9.0-windows10-x64-v7 and TensorFlow-gpu 1.9.0. The hardware configuration includes an Intel i7-9750H CPU and RTX series graphic cards. The training and testing steps based on TensorFlow can be summarized as follows: (1) Place different types of pictures into the corresponding folders and generate path index files for the training and validation sets. (2) Based on the path index files, transform the data to the tfrecords format to accelerate the data stream management process. (3) Build the network structure, set the hyperparameters, import tfrecords format data for training and validation, and save the models after the network converges to a stable state. (4) Extract the 2D-APM features of test signals and use the trained network to achieve accurate classification.

\section{OBSERVATION DATA} \label{sec:Observation}

The effective use of a deep learning classifier depends on the formulation of the scientific approach, which should be based on sufficient considerations of the availabilities of training data in terms of the quality and quantity. This paper aims to provide a method for identifying new pulsar candidates; hence, a more flexible way of training the network is needed to eliminate the limitations of the quality, quantity and types of existing observation data. One simulation method of TOAs based on the NHPP is adopted here to generate the 2D-APMs of the training set. The observation data of three pulsars from the RXTE are applied to generate the 2D-APMs of the test set. However, one uniform setting criterion of the time resolution of the periodic axis is proposed here to address a pulsar signal with an unknown period and unknown quality in practice.

\subsection{Training Set} \label{subsec:train set}

\subsubsection{TOA generation based on the Poisson distribution} \label{subsubsec:toas simulation}
Assuming that the TOA of X-ray photons obeys the NHPP, the probability of receiving $k$ photons in time $\Delta t$ should satisfy the following formula \citep{2011IEEETAES...47(4)...2317}:

\begin{equation}\label{equa3-1}
P(N_{\Delta t}=k)=\frac{(\lambda _{\Delta t})^k{\rm exp}(-\lambda _{\Delta t})}{k!}
\end{equation}
where $\lambda _{\Delta t}=\int_{\Delta t}\lambda(\xi)\,d\xi$, $\lambda(\xi)$ is the overall rate function, which represents the aggregate rate of all photons that arrive from the X-ray pulsar and 	 background.

Assume that $t_i$ is the TOA of the ith photon, $\Delta t=t_{i+1}-t_{i}$ is the time interval between the two photons, and $F(z)$ is the probability distribution function about $\Delta t$. The corresponding formula is expressed as follows:

\begin{equation}\label{equa3-2}
P(\Delta t>z)=1-F(z)
\end{equation}

Formula (\ref{equa3-1}) shows that $P(\Delta t>z|t_i=t)$ satisfies

\begin{equation}\label{equa3-3}
P(\Delta t>z)=P(N_{\Delta t}=0)={\rm exp}(-\int_{t_i}^{t_i+\Delta t}\lambda(\xi)\,d\xi)={\rm exp}(-(\Lambda(t_i+\Delta t)-\Lambda(t_i)))
\end{equation}
where $\Lambda (t_i)=\int_{0}^{t_i}\lambda(\xi)\,d\xi$. Combined with formula (\ref{equa3-2}), we can obtain

\begin{equation}\label{equa3-4}
t_{i+1}=t_i+\Delta t = \Lambda^{-1}(\Lambda(t_i)-{\rm ln}(1-F(z)))
\end{equation}

If the function $G$ is monotonically increasing, then $x=G^{-1}(U)$ is the only solution of $G(x)=U$. The TOAs can be generated by the inverse function method of random numbers based on the Monte Carlo approach.

Assume that $U$ is a random variable that obeys the uniform distribution from $0$ to $1$; then, $1-U$ satisfies the same probability distribution. Therefore, let $U=1-F(z)$, with which formula (\ref{equa3-4}) can be changed to

\begin{equation}\label{equa3-5}
t_{i+1}=t_i+\Delta t = \Lambda^{-1}(\Lambda(t_i)-{\rm ln}U) =\Lambda^{-1}(\Lambda(t_i)+E) 
\end{equation}
where $E$ is a random variable that obeys the exponential distribution with one parameter. The TOAs can be randomly generated by solving this recursive formula. Based on the difference method, the time complexity is linear, i.e., $O(n)$.

\subsubsection{Training data set production} \label{subsubsec:train set produc}

The characteristics and quality of X-ray pulsar signals are determined by the pulsar rotation period, pulse shape, number of arriving photons and emission flux. Considering that the 2D-APM, which is obtained based on epoch folding, has no period information, the rotation period $T_p$ of the simulation signal is uniformly set to 100 ms in the process of devising the training set. According to Section \ref{subsec:2dapm}, when the pulsar profile is simulated with the double Gaussian components, the most substantial factor that affects the characteristic distribution of 2D-APM is the peak separation. Therefore, we only consider the peak separation instead of the pulse width and amplitude ratio of the double peaks in the simulation of the pulsar profile. The amplitude ratio of the double peaks is set to 1:1, and the standard deviations of the double Gaussian components are set to 0.04. The signal quality is mainly determined by the number of arriving photons and emission flux. For the convenience of simulation, the quality of 2D-APM is only controlled by adjusting the number of arriving photons, and the emission flux remains unchanged. The number of photons from an X-ray pulsar and the cosmic background in a single period are set to 50 and 350, respectively. The parameters that need to be controlled in the simulation are the peak separation and number of arriving photons.

To ensure that the 2D-APM has rich time-domain information, the time resolution of its periodic axis should be set to a reasonable value. If the time resolution is too large, then the effective information in the entire image will be insufficient. Inversely, a time resolution that is too small reduces the contrast in the entire image, which causes a lack of detailed information regarding the profile changes. In addition, the longer the observation duration is and the shorter the folding period is, the larger the impact of the period error on the pulsar profile is. Considering that the time resolution of the periodic axis is relative to the candidate period, one empirical formula is proposed here to confirm the periodic resolution:

\begin{equation}\label{equa3-6}
\frac{RL}{{T_c}^2}=C
\end{equation}
where $R$ is the time resolution of the periodic axis, $L$ is the signal length, $T_c$ is the candidate period and $C$ is a constant value (set to 3 here). In this way, as the observation time or candidate period changes, the time resolution of the periodic axis also adjusts accordingly to ensure that the feature cover ranges of the positive samples of different data sets are equivalent.

\begin{figure}
	\centering
	\includegraphics[width=14cm,height=18cm]{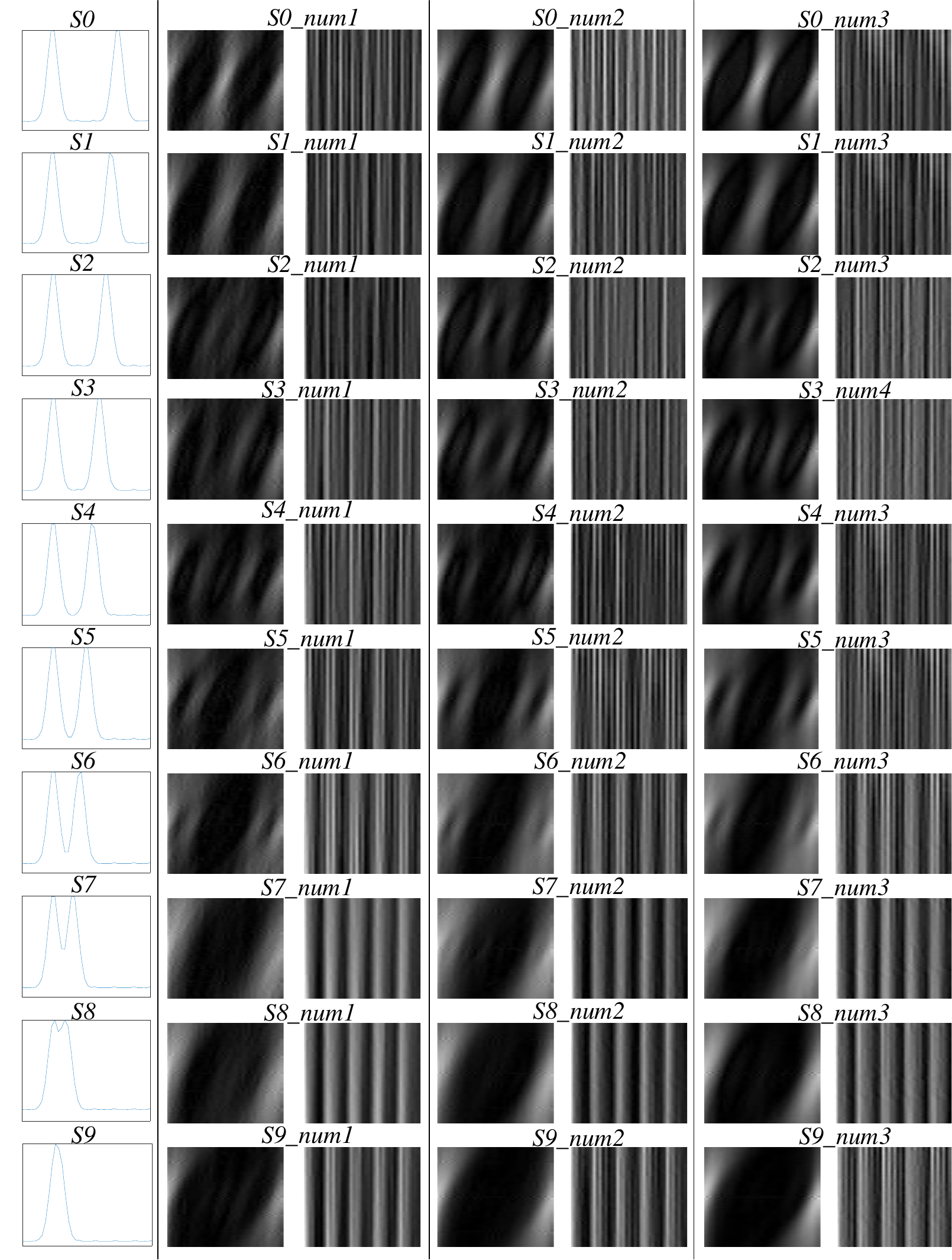}
	\caption{The pulsar profiles and samples of 30 data sets in the training set. The leftmost column shows the 10 profiles of the training set, and the number appears at the top of each image. Each row displays positive and negative samples with different numbers of photons of the same pulsar profile. The left image of each data set is a positive sample, and the right image of each data set is a negative sample. The vertical axis of each sample represents the folding period; the horizontal axis represents the phase; and the name of the data set appears above each group of samples.}
	\label{fig:repre of the train set }
\end{figure}

According to the previously mentioned criteria, ten pulsar profiles named $S0 (0.5)$, $S1 (0.45)$, $S2 (0.4)$, $S3 (0.35)$, $S4 (0.3)$, $S5 (0.25)$, $S6 (0.2)$, $S7 (0.15)$, $S8 (0.1)$ and $S9 (0.05)$ are designed. The number in the brackets represents the peak separation with a normalized phase range of $0\sim1$. The number of photons employed in the simulation are $N1 (5000)$, $N2 (10000)$ and $N3 (20000)$, and the number in the brackets represents the specific number of photons. Additionally, $30$ combinations of the data set are obtained from $S0\_N1$ to $S9\_N3$.

The period search range of this experiment is set between $10~ms$ and $1~s$, which corresponds to 1 Hz$\sim$100 Hz in the frequency domain, which can cover most normal pulsars. When the space telescope is aimed at the pulsar, it can receive X-ray photons from both the pulsar and the universe, which is the background. When the photons are processed with the correct period, we can extract the 2D-APM features, which are positive samples. However, when using other periods to exact the features, we can only obtain irregular samples, which are negative samples, just as the lost target situation, which received random photons from the background. Therefore, the negative samples are constructed by generating 2D-APMs around the frequency point different from the rotation frequency, while the positive samples are constructed by generating 2D-APMs near the rotation frequency. There are very few effective frequency points compared with the whole searching range, which often leads to an imbalance in the proportion of positive and negative samples. When the number of negative samples is considerably greater than the number of positive samples, the network can hardly learn the effective information of the positive samples. As the 2D-APM covers a wide range of time-domain information, the 2D-APMs that contain the effective frequency points are classified as positive samples, and the other samples are classified as negative samples. Dense sampling is conducted near the rotation frequency, and sparse uniform sampling is conducted in the other frequency domains to ensure a balance between positive samples and negative samples. Each data set generates 500 positive samples and 500 negative samples. In this way, the total size of the training set is 30,000.

Figure \ref{fig:repre of the train set } shows the randomly sampled result of the training set. Most of the negative samples show an irregular vertical line arrangement. There also exist some negative samples with more disordered patterns. It can be easily understood that an irregular noise signal will be produced when the period signal is folded according to the inaccurate period. The positive samples show clearer features and less grainy noise from left to right in each row, as the quality of the features obtained by folding improves gradually as the photon numbers of the corresponding data set increase. The number of positive samples is expanded by dense sampling near the rotation period. However, only one positive example of each data set is randomly shown in the figure; the remaining examples are similar to the upward and downward translation of the example, which can be understood to some extent as an image transition of the data enhancement methods. The positive samples from top to bottom in each column show regular changes as the peak separation decreases. At the level of graphic expression, the change law is similar to a shear wave in the centre conducting to both sides. By learning the discrete change law of 10 kinds of pulsar signals, the network is expected to possess the generalization ability to identify more types of pulsars. All the positive samples clearly show consistent feature coverage and slant degrees of the bright spots, which reveal the rationality and applicability of the setting criteria of the periodic resolution.

\subsection{Test Set} \label{test set}

\subsubsection{RXTE data extraction \label{subsec:data extraction}}

The low earth satellite RXTE was decommissioned in January 2012. During its 16-year service, a large number of valuable X-ray pulsar data were observed and published, which greatly advanced pulsar timing observation research. The data obtained by the RXTE, which are stored in Flexible Image Transport System (FITS) format, can be accessed from the HEASARC by web-based browsing or an anonymous file transfer protocol (FTP).

From the characteristic shown in Section \ref{subsec:2dapm}, we know that the distribution law of the 2D-APM is directly related to the peak separation of the pulse profile. Therefore, we selected PSR B0531+21, PSR B0540-69 and PSR B1509-58 (in a large-to-small double-peak distance sequence) as experimental pulsars. Their related parameters are listed in Table \ref{tab:selec pulsa para}. The observation programs of these three pulsars in the RXTE database are P96802, P10206 and P50705.

The proportional counter array (PCA) is one of the main RXTE detectors. The PCA is composed of five proportional counter units (PCUS) and has a total collection area of approximately 6500 $cm^2$ and an effective energy range of 2$\sim$60 keV. Its best time resolution is 1 us, and its best energy resolution is 18$\%$ at 6 keV. The two PCA data modes are the Science Array-Mode and Science Event-Mode. The Array-Mode data record the photon numbers of each channel at equal time intervals, and the Event-Mode data record the arrival time and energy band of each photon. The Event-Mode data utilized in the paper are applied in the Good Xenon format.

We used Heasoft software for RXTE data preprocessing as follows: (1) Use the XDF tool visual interface to identify and record the file index of the data to be processed. (2) Filter the observation task to generate suitable time intervals with Xtefilt and Maketime. (3) Use the Fselect command to filter clock events. (4) Use the Faxbary command to correct the TOAs, and then use the Seextrct command to extract the light curves that contain the photon data of all energy bands. (5) Use the Efsearch tool to confirm the optimal period of each observation task. Save the optimal period and light curve information as a series of txt files for further processing.

\subsubsection{Test data set production} \label{subsubsec:test set produc}

\begin{figure}
	\centering
	\includegraphics[width=15cm,height=12cm]{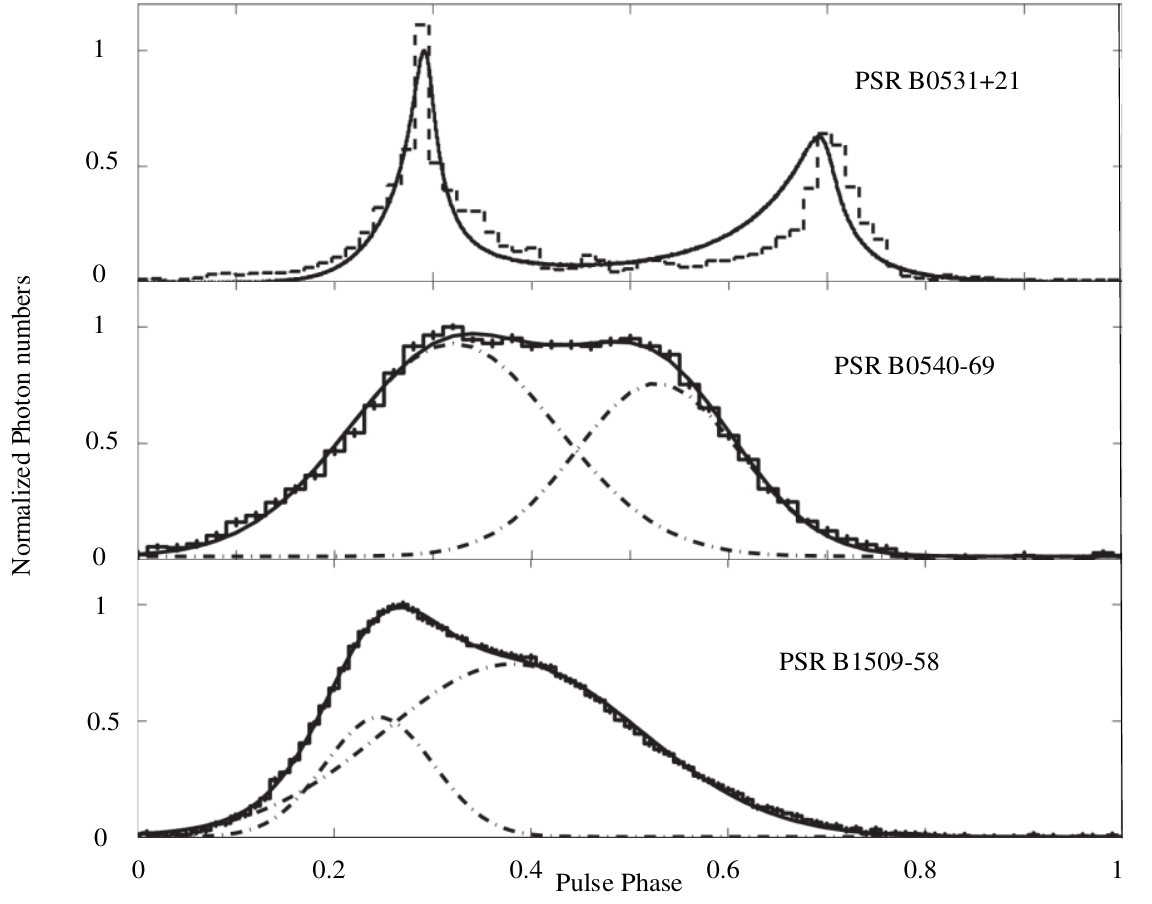}
	\caption{X-ray profiles of PSRs B0531+21, B0540-69, B1509-58 with RXTE. The step curve records the actual photon numbers of each phase bin, the solid line represents the fitted profile, and the dashed line represents the double Gaussian components of the fitted profile. The figure shows the results from \citet{2012APJS...199(2)...32}.}
	\label{fig:X-ray profiles of PSRs}
\end{figure}

The profile shapes of the three pulsars are shown in Figure \ref{fig:X-ray profiles of PSRs}. The peak separations of B0531+21, B0540-69 and B1509-58 decrease to $\sim0.4$, 0.2$\sim$0.25 and 0.15$\sim$0.2, respectively. Their pulse widths and amplitude ratios of double peaks are different. In general, they are quite different from the 10 kinds of simulation profiles. If the trained network can successfully identify them, then it can show the generalization ability of the training model and the universality of the proposed method. Among these three X-ray pulsars, B0531+21 is the brightest pulsar, and B0540-69 is the darkest pulsar. To verify the anti-noise performance and robustness of the network, three data sets with different observation durations are prepared for each pulsar, as shown in Table \ref{tab:para of sample sets}. Robustness refers to the ability of the network to adapt to 2D-APM instances of different quality. For the test set, 300 positive samples and 300 negative samples were created for each of the 9 types of data sets shown in Table 2; the total size of the test set is 5400. The generation method of positive samples and negative samples is the same as that described in Section \ref{subsubsec:train set produc}. 

\begin{figure}
	\centering
	\includegraphics[width=16cm,height=8cm]{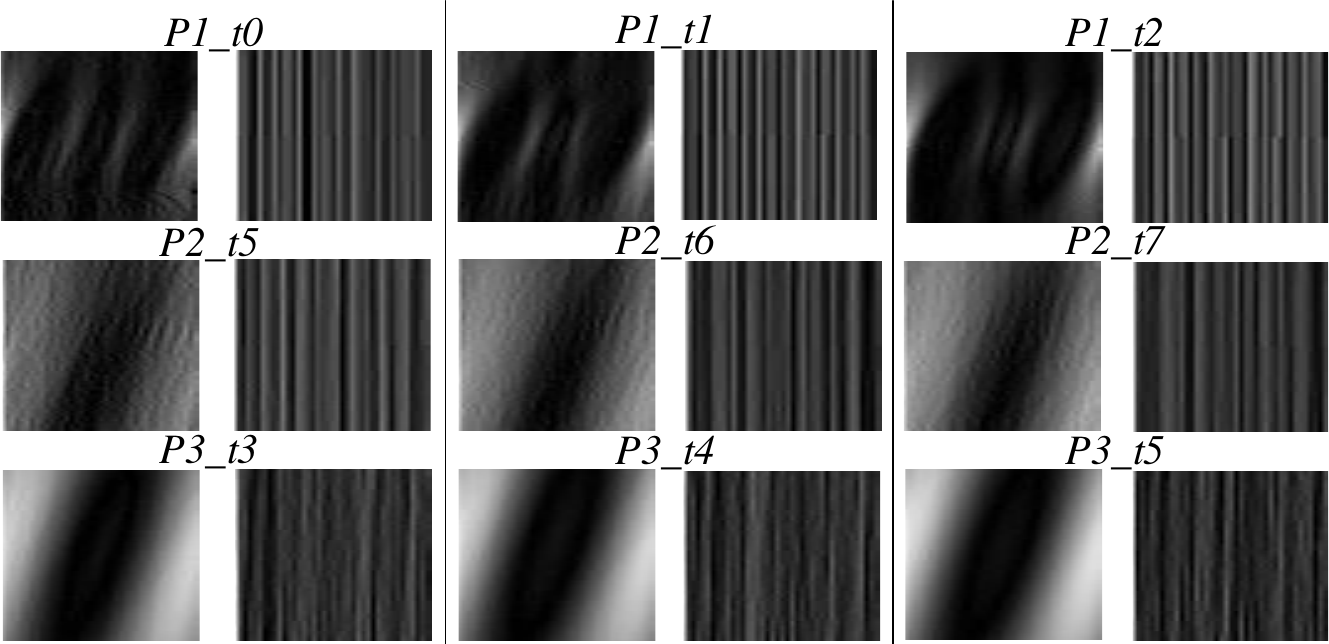}
	\caption{The samples of 9 data sets in the test set. Each row displays positive and negative samples with different observation durations for the same pulsar. The left image of each data set is a positive sample, and the right image is a negative sample. The vertical axis of each sample represents the folding period, the horizontal axis represents the phase, and the name of the data set appears above each group of samples.}
	\label{fig:representation of test sets}
\end{figure}

A randomly sampled example of the test set is shown in Figure \ref{fig:representation of test sets}. Similar to Figure \ref{fig:repre of the train set }, most of the negative samples show irregular patterns, and the positive samples show clearer features and less grainy noise from left to right in each row. Compared with the differences in the pulsar profiles between the simulation data and the observation data, the 2D-APMs of the three pulsars show the distributions of characteristics and change rules similar to those of the simulation data, such as the $P1$ series of positive samples and $S2$ series of positive samples. Notably, although the observation time and period of the three pulsar signals are different, their positive samples also show the same feature coverage under the guidance of the reasonable periodic resolution obtained by formula \ref{equa3-6}.

\begin{table}[!htbp]
	\centering
	\caption{Parameters for the three pulsars.\label{tab:selec pulsa para}}
	\resizebox{110mm}{15mm}{
		\begin{tabular}{ccccc}
			\hline
			Number& Name &Pulsar period&Binary&Source total flux\\
			{} & {} & s &{} &$\rm ph\,cm^{-2}s^{-1}$ \\
			\hline
			1 & B0531+21 & 0.03308 &No & 3200 \\
			2 & B0540-69 & 0.05035 &No & 160\\
			3 & B1509-58 & 0.15065 &No & 180\\
			\hline
	\end{tabular}}
\end{table}

In the actual searching task, 2D-APM samples are obtained by a sliding window, and overlap occurs between adjacent samples. Compared with the traditional epoch folding algorithm, the number of samples is identical. Therefore, the increase in computational complexity here is linear, which mainly depends on the feature extension dimensions along the periodic axis. In this experiment, the feature dimensions along the periodic axis are $299$; hence, the computational complexity in the feature extraction process is $299$ times that of the traditional epoch folding algorithm.

\begin{table}[!htbp]
	\centering
	\caption{Parameters of the sample sets.\label{tab:para of sample sets}}
	\resizebox{80mm}{30mm}{
		\begin{tabular}{cccc}
			\hline
			ID& Sample Set & Pulsar & Data Length\\
			\hline
			1& $P1\_t0$& B0531+21& 1000(1 s) \\
			2& $P1\_t1$& B0531+21& 2000(2 s)\\
			3& $P1\_t2$& B0531+21& 4000(4 s)\\
			4& $P2\_t5$& B0540-69& 400000(400 s)\\
			5& $P2\_t6$& B0540-69& 600000(600 s)\\
			6& $P2\_t7$& B0540-69& 800000(800 s)\\
			7& $P3\_t3$& B1509-58& 200000(200 s)\\
			8& $P3\_t4$& B1509-58& 300000(300 s)\\
			9& $P3\_t5$& B1509-58& 400000(400 s)\\
			\hline
	\end{tabular}}
\end{table}

\section{RESULTS} \label{sec:results}

\subsection{Network Training Results \label{subsec:Network Training}}

\begin{figure}
	\centering
	\includegraphics[width=12cm,height=7.5cm]{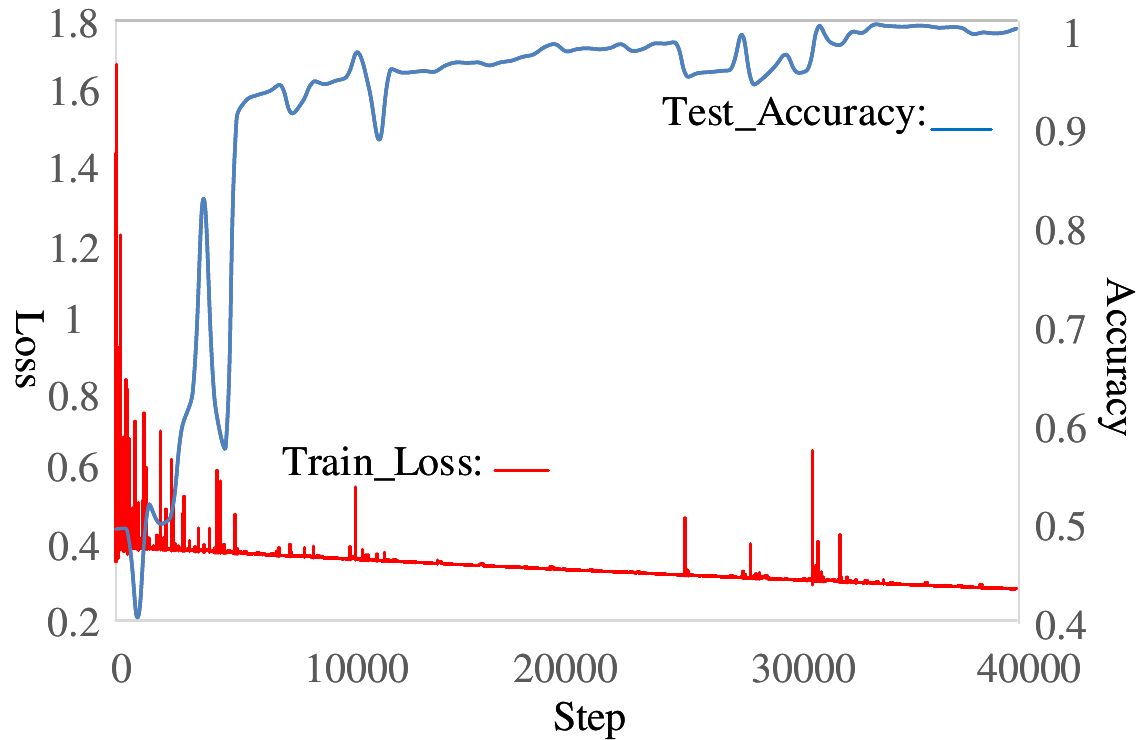}
	\caption{The error loss curve of the training set is shown in red, and the classification accuracy curve of the test set is shown in blue. The horizontal axis represents the training steps. One step in the training process involves randomly selecting sample sets of the predetermined batch size for training, while one step in the test process involves testing all 5400 samples in the test set. The vertical axis on the left represents the training loss, and the vertical axis on the right represents the classification accuracy.}
	\label{fig:train loss}
\end{figure}

We input the sample sets into the network and show the recorded training process in Figure \ref{fig:train loss}. In the initial 2,000 steps of the training process, the classification accuracy on the test set remains steady at approximately 50\%, although the training error decreases gradually, which shows that the network has not yet learned the key classification characteristics. In the initial stage, the training error oscillates greatly; it is understandable that there exist large differences among the samples of the training set, and the network is still looking for the correct path to reduce the total error of the whole training set. The classification accuracy on the test set rises quickly above 90\% in approximately 5000 steps, which shows that the network learning process is on track. The network training error decreases continuously, but its classification accuracy remains unchanged when it increases to 99\%, which indicates that overfitting may have occurred. At this time, we terminate the training and save the model. Generally, the learning process is relatively fast; it is directly related to the appropriate learning rate and batch size but cannot be separated from the reduction in network size. 

\begin{figure}
	\centering
	\includegraphics[width=12cm,height=7.5cm]{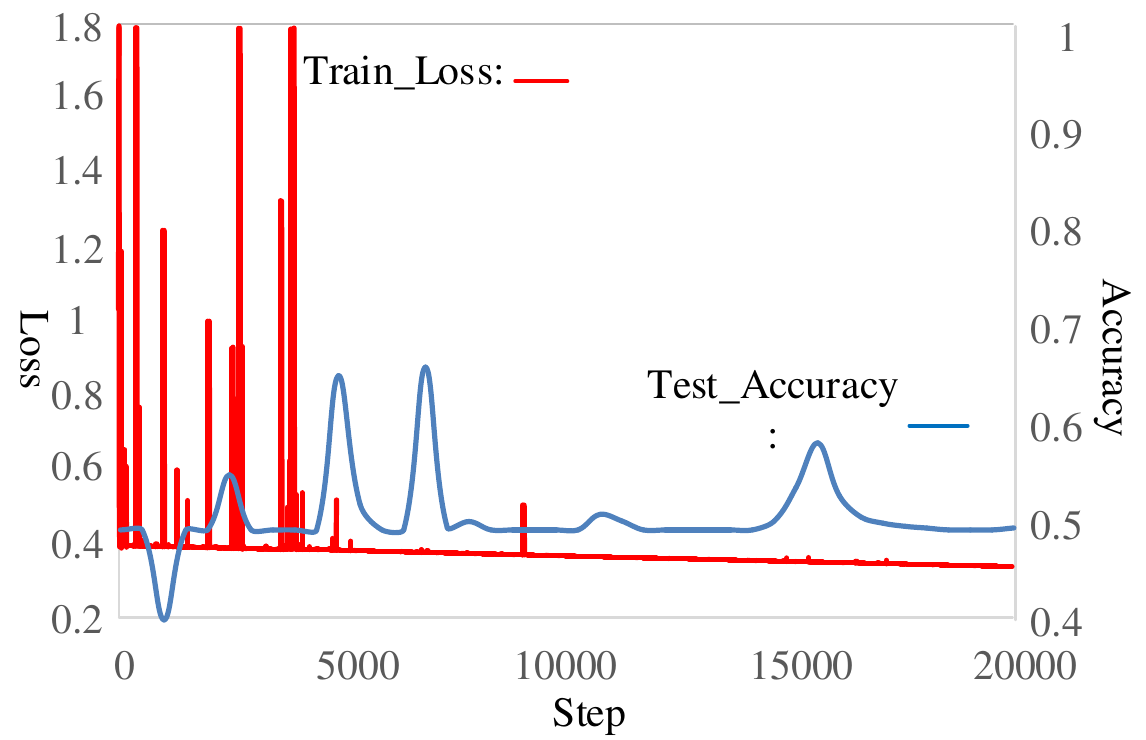}
	\caption{Similar to Figure \ref{fig:train loss}, the error loss curve of the training set is shown in red, and the classification accuracy curve of the test set is shown in blue. }
	\label{fig:train loss2}
\end{figure}

To illustrate the generalization power of the network, the training set with only the $S0$ waveform is used to train another network; the training process is shown in Figure \ref{fig:train loss2}. Figure \ref{fig:train loss} and Figure \ref{fig:train loss2} are identical in size to the S0 waveform training set. Figure \ref{fig:train loss} illustrates that the network has learned the features of 10 kinds of waveforms and their generalized features. Therefore, after the training error curve of Figure \ref{fig:train loss2} converges, the network can learn the feature of the single waveform. However, the classification accuracy of the test set does not improve since the training error decreases continuously, which shows that a training set with only one waveform cannot guide the network to achieve the generalization ability.

Some additional explanations are given as follows: (1) Compared with the reference value 0.4, the training error varies at a relatively low level due to the effect of the regularization term of the loss function, which accelerates the training process and reduces the likelihood of overfitting to some extent. (2) The total error of the test set is inversely proportional to its classification accuracy. To show the training results more clearly, we show the classification accuracy curve instead of the error curve. (3) During training, some oscillations occur that affect the overall learning process. These oscillations are largely due to interference from negative samples without acceptable characteristics. These samples may cause the network to become trapped in a local optimum, which is the key reason why the momentum optimizer is chosen here. (4) The network does not reach its performance limit; hence, its performance can still be improved by further expanding the sample sets with additional types of data sets and increasing the training time.

\begin{figure}
	\centering
	\includegraphics[width=12cm,height=12cm]{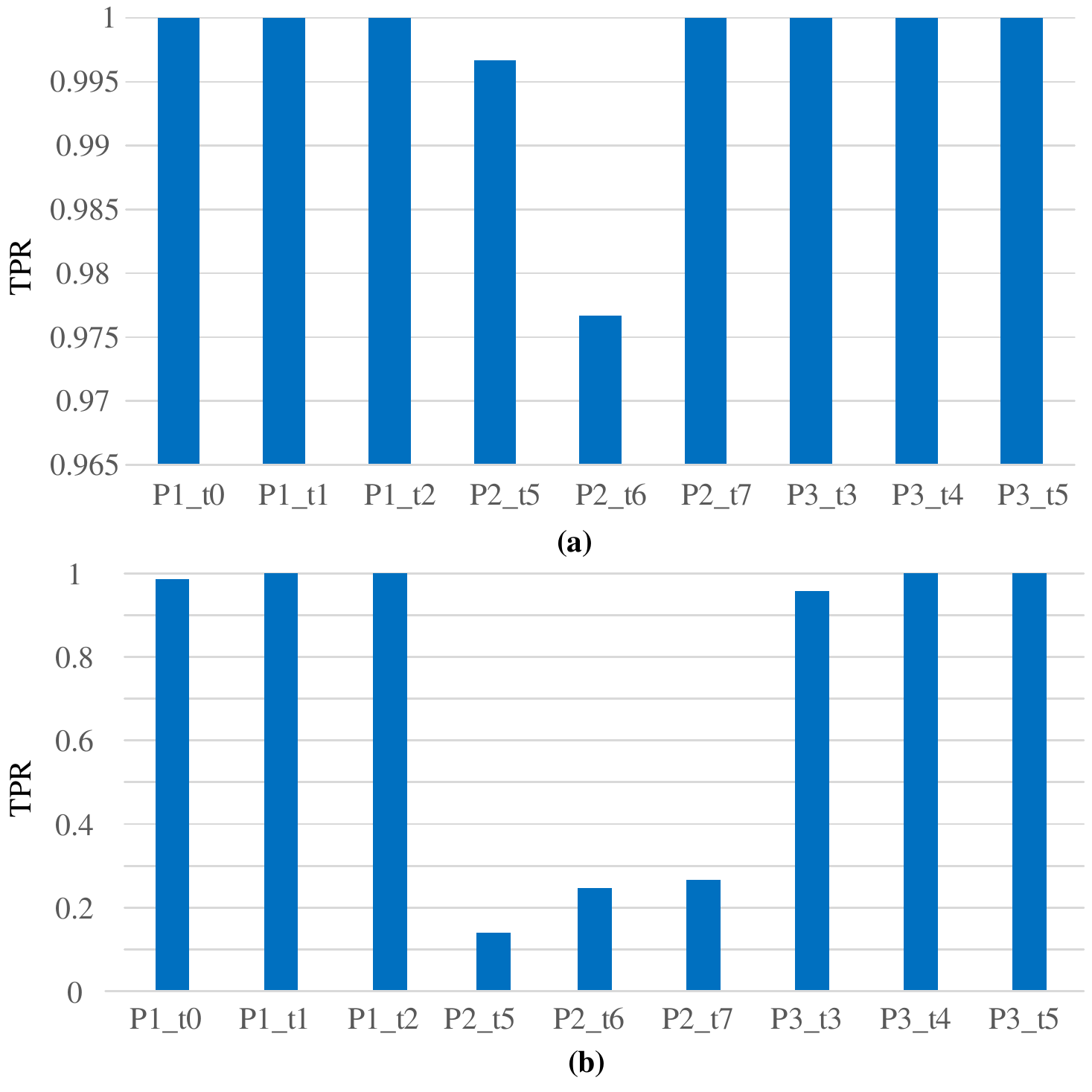}
	\caption{Network classification results (a) and SVM (based on HOG) classification results (b) of the 9 types of positive sample sets shown in Table \ref{tab:para of sample sets}.}
	\label{fig:classification results}
\end{figure}

\subsection{Analysis of the Results and Discussion \label{subsec:analysis and discussion}}

%textcolor{blue}{}
%textcolor{red}{\sout{}}

In the pulsar sample classification task, the true positive rate (TPR) and false positive rate (FPR) are usually employed to evaluate an algorithm's performance. The TPR indicates the probability or proportion of correct matches in positive samples, and the FPR indicates the probability or proportion of incorrect matches in negative samples. A higher TPR and a lower FPR indicate better algorithm performance. In our experiment, among 2700 negative samples, there are 2 samples divided into positive results, which corresponds to a very low FPR. The network's classification results for the positive samples of the test set are presented in Figure \ref{fig:classification results} (a), which shows a high TPR in classifying the 9 types of datasets. For $P1$ and $P3$, their signal quality is acceptable; thus, the network has minimal pressure to distinguish them. Among these three X-ray pulsars, $P2$ is the darkest. Combined with Figure \ref{fig:representation of test sets} in section \ref{subsubsec:test set produc}, it can be seen that the positive samples of $P2$ contain the most noise compared with those of $P1$ and $P3$. Therefore, when using $P2\_t5$ and $P2\_t6$ positive sample sets with relatively short observation duration times, the network accuracy is slightly lower.

Generally, the feature mining effect of the neural network is better than the classification results of the specific rules or traditional machine learning methods, especially for the case where a large amount of training data can be obtained. To objectively illustrate the advantages of the DCNN, a support vector Machine (SVM) algorithm based on a histogram of oriented gradient (HOG) is used to classify the positive and negative samples as a reference algorithm\citep{2005CVPR...886}. The specific configuration includes 9 orientation bins, 32$\times$32 pixel cells, 2$\times$2 pixel blocks of the HOG and a linear kernel of the SVM. In the experiment, among 2700 negative samples, there are several samples divided into positive result, which also corresponds to a very low FPR. Figure \ref{fig:classification results} (b) shows the network's classification result on the positive samples. Compared with Figure \ref{fig:classification results} (a), it can been seen that the SVM also has minimal pressure to distinguish between P1 data sets and P3 data sets. For P2 data sets, which contain much more noise, the SVM performs quite poorly. Generally, the SVM algorithm has better mathematical interpretablity and performs well when addressing small data sets with satisfactory quality and long observation times. In practical scenarios with data sets with different qualities, larger sizes and more complex features, the DCNN has better applicability.

\begin{figure}
	\centering
	\includegraphics[width=15cm,height=7.5cm]{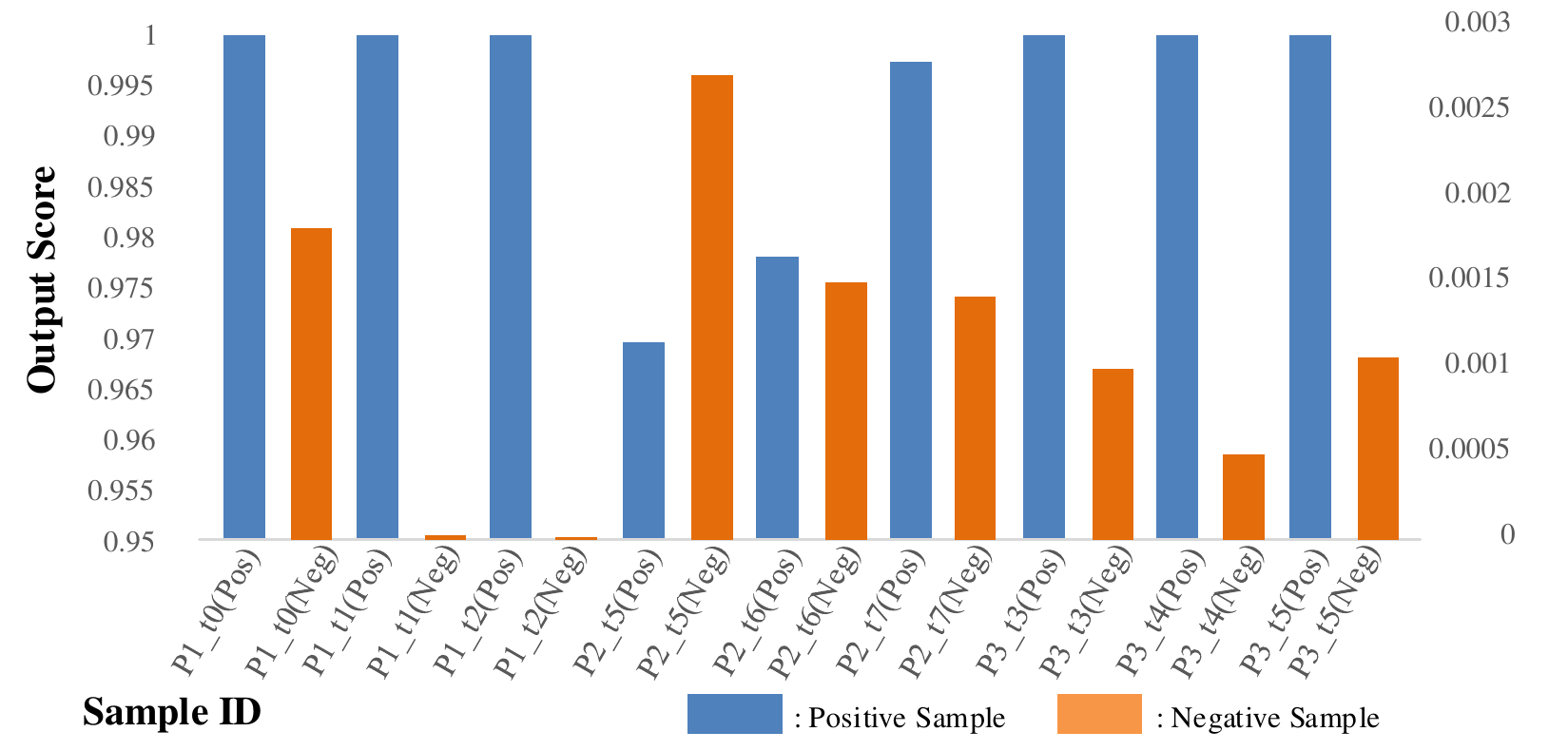}
	\caption{
		Histograms of the output scores for different test sets. Here, the output score refers to the average score of the output `yes'. The final layer of the network is the softmax layer, so the output score ranges from 0 to 1. The horizontal axis lists the IDs of positive and negative samples from different datasets. The blue column represents the score of the positive samples, and its corresponding vertical axis is shown on the left. The yellow column represents the scores of the negative samples, and its corresponding vertical axis is shown on the right. }
	\label{fig:output score}
\end{figure}

To analyse the network performance more specifically, the output scores of the final layer for different datasets are shown in Figure \ref{fig:output score}. The output distributions of the positive and negative samples are significantly different and do not overlap, which further shows the superior classification ability of the network. The output scores of the negative samples are very small, and there is minimal relation between the negative samples of different datasets. For the same pulsar, the longer the observation time is and the clearer the characteristic distribution of the 2D-APM is, the higher the output score will be. Therefore, the output score of the $P2$ positive samples increases as the observation duration time increases. 
As an increasing number of samples are introduced, the results in Figure \ref{fig:classification results} (a) will be closer to the trend shown in Figure \ref{fig:output score}. The TPR of $P2\_t5$ in Figure \ref{fig:classification results} (a) is higher than we expected, which reflects the randomness of the data extraction procedure.

The number of positive samples, which correspond to a few points in the whole spectrum, are far fewer than the negative samples. In the actual scene, the requirement of the FPR is higher than that of the TPR. However, the output scores of these samples are less than 0.7. Therefore, these false positive samples can be filtered by setting thresholds. The samples that are classified as effective signals can be verified multiple times with the data of the non-repetition period to further reduce the FPR and enhance the practicability of the method. Note that these strategies have minimal impact on the TPR.

\begin{figure}
	\centering
	\includegraphics[width=14cm,height=6cm]{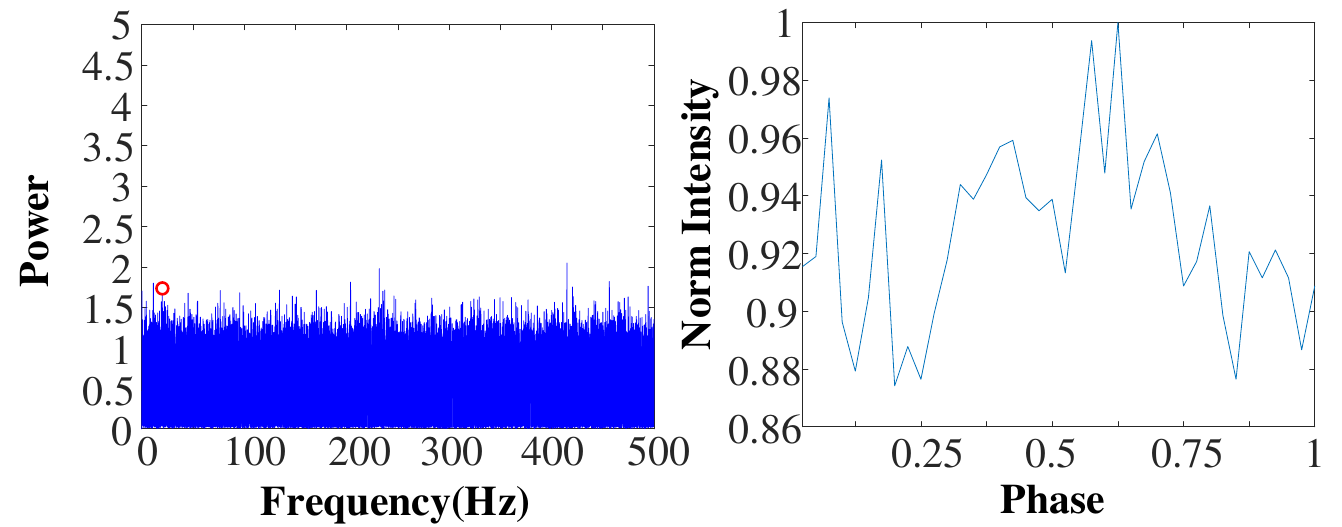}
	\caption{
		The frequency spectrum and the corresponding traditional folding profile of B0540-69 when the observation time is 400 s. The rotation frequency is indicated by the red circle in the left picture.  }
	\label{fig:traditional result}
\end{figure}

Figure \ref{fig:traditional result} shows the frequency spectrum and traditional folding profile of B0540-69. When the observation time is 400 s, the self-rotation frequency points that we need to detect are indicated by the red circle, which is almost submerged in noise. The corresponding folding profile does not have obvious information about the shape and phase. Therefore, pulsar candidate searches are very difficult to perform. Comparing the classification results in Figure \ref{fig:classification results} (a) with the clear features of the $P2\_t5$ positive sample in Figure \ref{fig:representation of test sets}, the effectiveness of the proposed feature extraction method in pulsar searching can be verified.

\begin{figure}
	\centering
	\includegraphics[width=14cm,height=8cm]{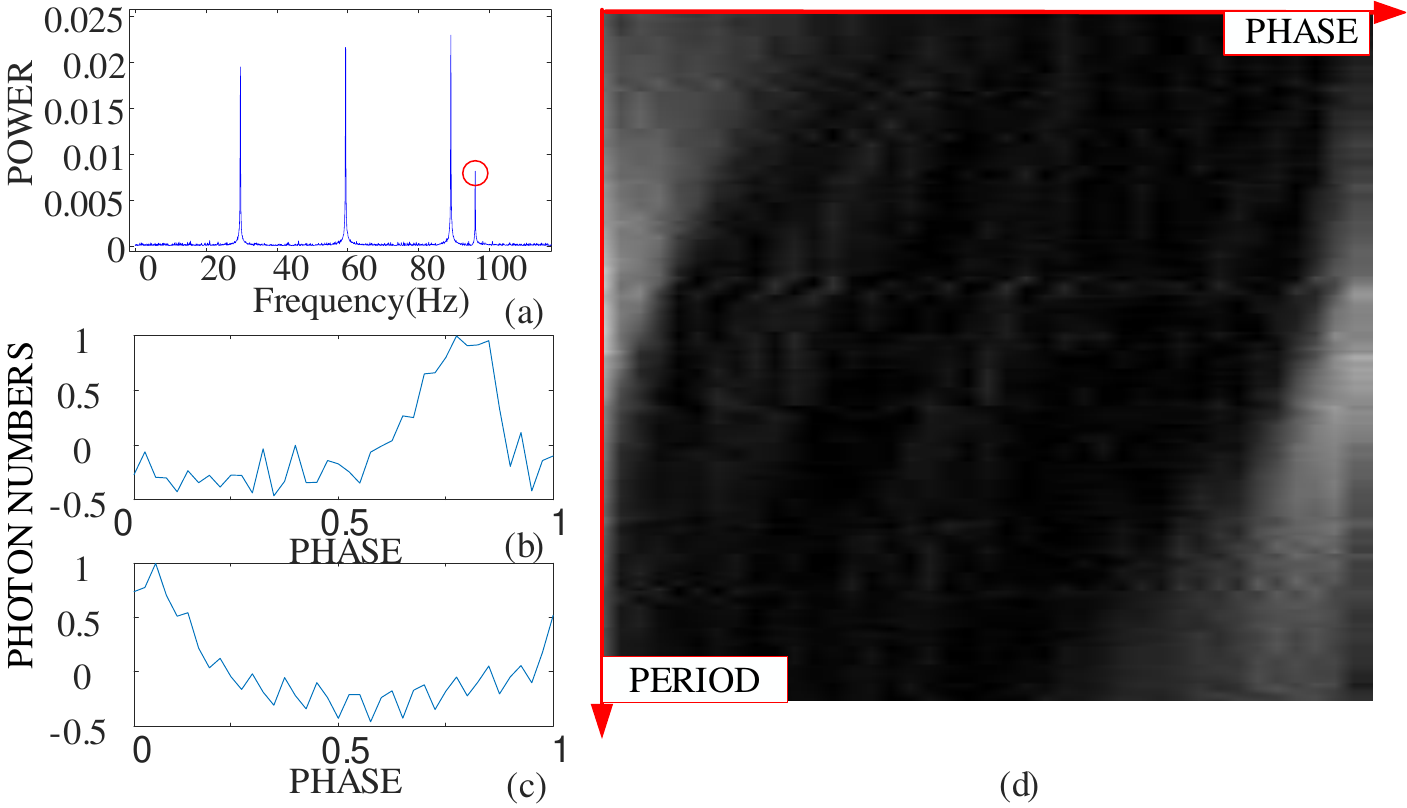}
	\caption{Frequency spectrum of the original signal of B0531+21 in (b). The interference with a fundamental frequency of $\sim$96 Hz is marked with a red circle. At the interference frequency, the traditional profile, autocorrelation profile and 2D-APM are shown in (b), (c) and (d), respectively.}
	\label{fig:False positive sample}
\end{figure}

Unlike radio observations, X-ray observations are not often affected by statistical correlation interference, such as observed in telecommunication signals. However, we note some obvious statistical correlation interference at $\sim$96 Hz of B0531+21. Four peaks exist in the frequency spectrum of Figure \ref{fig:False positive sample}(a). The first three peaks are the rotation frequency and harmonic frequencies of B0531+21, while the fourth peak likely originates from the instrument. However, the characteristics expressed by the interference point, including its traditional folding profile (refer to Figure \ref{fig:False positive sample}(b)), the autocorrelation profile (refer to Figure \ref{fig:False positive sample}(c)) and the 2D-APM (refer to Figure \ref{fig:False positive sample}(d)), are very similar to those of the effective signal. Thus, a more targeted method is needed to identify them.

From the point of energy, the three pulsars employed in the experiment are actually bright. However, they are representative of the point of the wave form and periodic laws of normal pulsars. Moreover, for weak pulsar signals, a longer observation time is needed to obtain a sufficient number of photons. The long observation period is needed to address more difficulties that affect the analysis, such as the attenuation of the rotation speed, the Doppler frequency shift caused by the change in distance between the spacecraft and the pulsar, and even the frequency correction and phase alignment of multiple discontinuous observation data \citep{2017APJ...845(2)}. These difficulties will undoubtedly introduce more work. We know that extending the observation period is very necessary in practical application, but our purpose is to propose a new effective method in the current stage. We have constructed data sets with different observation times to simulate the lack of features caused by an insufficient number of photons. An analysis of weaker pulsar signals is our next research direction.

\section{CONCLUSIONS} \label{sec:conclusions}

In this paper, we proposed a new classification method for pulsar candidates, including a new feature modelling method and combination application using a deep convolutional neural network. At the feature representation level, the proposed feature model has an unbiased phase, a consistent distribution, rich information and a strong anti-noise ability. At the feature perception level, we selected Inception-ResNet, which has a strong expressive ability, to learn the 2D spatial information of a feature. Moreover, we proposed a uniform setting criterion regarding the periodic resolution and a simulation framework for training the network, which guarantee the feature consistency of the pulsar candidates, training flexibility and generalization power of the network, which enhances the practicability of the overall method.
The experimental results, which are based on both simulated signals and RXTE observation data, demonstrate the superiority of the modelling method and trained deep convolutional neural network in mining model information. The network in this paper has not reached its performance limit and could achieve better performance in practical applications if the sample sets are expanded, including additional observation durations and types of simulation profiles. Although the proposed method was specifically developed for X-ray pulsars and is suitable for performing pulsar searches at X-ray energies, its modelling and machine learning methods could easily be transplanted to the radio field.

\begin{acknowledgements}
This work was funded by the National Natural Science Foundation of China (NSFC)
under No. 11973021.
\end{acknowledgements}

\label{lastpage}
\end{document}